\documentclass[a4paper,10pt,twoside,linkcolor=blue,anchorcolor=blue,citecolor=blue,urlcolor=blue]{cpc-hepnp}
\usepackage{CJK,upgreek,fancyhdr}
\usepackage{multicol}
\usepackage{graphicx}
\usepackage{booktabs}
\usepackage{amssymb,bm,mathrsfs,bbm,amscd}
\usepackage[tbtags]{amsmath}
\usepackage{lastpage}
\usepackage[figureright]{rotating}
\usepackage{color}
\usepackage{ulem}
\usepackage{threeparttable}
\usepackage{verbatim}
\usepackage{textcomp}
\usepackage{gensymb}
\usepackage{subfigure}
\usepackage{multirow,booktabs}
\usepackage{float}
\usepackage{orcidlink}
\usepackage{tikz}
\hypersetup{
colorlinks=true,
linkcolor=black,
citecolor=black
}

\begin{document}
\begin{CJK*}{UTF8}{gbsn}

\fancyhead[c]{\small Chinese Physics C~~~Vol. xx, No. x (202x) xxxxxx}
\fancyfoot[C]{\small 010201-\thepage}

\footnotetext[0]{Received 11 September 2025}

\newcommand{\orcidauthorA}{0000-0001-8938-9152}  


\title{Fragmentation of neutron-rich carbon isotopes on light targets at 27.5 MeV/nucleon
\thanks{Supported by the National Key R\&D Program of China (Contract No. 2023YFA1606403, 2023YFE0101600 and 2022YFA1605100) and the National Natural Science Foundation of China (Contract Nos. 12350007, 12275006, 12275007 and 12027809).}}

\author{ Zi-Yao Hu (胡梓瑶)$^{1}$ 
\quad Yan-Lin Ye (叶沿林)$^{~\orcidlink{0000-0001-8938-9152}}$$^{1;1)}$\email{yeyl@pku.edu.cn}
\quad Jian-Ling Lou (楼建玲)$^{~\orcidlink{0000-0002-8246-4807}}$$^{1}$
\quad Zai-Hong Yang (杨再宏)$^{~\orcidlink{0000-0001-5596-362X}}$$^{1}$\\
\quad Xiao-Fei Yang (杨晓菲)$^{~\orcidlink{0000-0002-1633-4000}}$$^{1}$
\quad Li-Sheng Yang (阳黎升)$^{1}$
\quad Wei-Liang Pu(蒲伟良)$^{1}$
\quad Kang Wei (魏康)$^{~\orcidlink{0000-0002-5120-5837}}$$^{1}$\\
\quad Ying Chen (陈莹)$^{1}$
\quad Hong-Yu Zhu (朱宏渝)$^{1}$
\quad Bo-Long Xia (夏博龙)$^{1}$
\quad Jia-Xing Han (韩家兴)$^{1}$
\quad Jia-Hao Chen (陈家豪)$^{1,3}$\\
\quad Kai Ma (马凯)$^{1}$
\quad Dong-Xi Wang (王东玺)$^{1}$
\quad Hao-Yu Ge (葛浩煜)$^{1}$
\quad Wen-Wu Wan (万文武)$^{1}$\\
\quad Jia-Wei Bian (边佳伟)$^{1}$
\quad Ze-Yu Du (杜泽宇)$^{1}$
\quad Zhe-Yang Lin (林\hbox{\scalebox{0.8}[0.96]{吉}\kern-.2em\scalebox{0.8}[0.96]{吉}}阳)$^{1}$
\quad Qi-Te Li (李奇特)$^{~\orcidlink{0009-0005-1319-550X}}$$^{1}$\\
\quad Zhi-Huan Li (李智焕)$^{1}$
\quad Ong-Hooi Jin (王惠仁)$^{~\orcidlink{0000-0002-7291-7809}}$$^{2}$
\quad Yan-Yun Yang (杨彦云)$^{~\orcidlink{0000-0002-5982-1706}}$$^{2}$
\quad Shi-Wei Xu (许世伟)$^{2}$\\
\quad Jun-Bing Ma (马军兵)$^{2}$
\quad Zhen Bai (白真)$^{2}$
\quad Kang Wang (王康)$^{~\orcidlink{0000-0003-1707-5490}}$$^{2}$
\quad Fang-Fang Duan (段芳芳)$^{~\orcidlink{0009-0003-7895-5149}}$$^{2}$\\
\quad He-Run Yang (杨贺润)$^{2}$
\quad Peng Ma (马朋)$^{2}$
\quad Xiang-Lun Wei (魏向伦)$^{2}$
\quad Tian-Li Qiu (邱天力)$^{2}$\\
}

\maketitle

\address{%
$^1$ School of Physics and State Key Laboratory of Nuclear Physics and Technology, Peking University, Beijing 100871, China\\
$^2$ Institute of Modern Physics, Chinese Academy of Sciences, Lanzhou 730000, China\\
$^3$ Institute of Applied Physics and Computational Mathematics, Beijing 100094, China\\
}

\begin{abstract}
Experimental and theoretical investigation of the fragmentation reaction in Fermi-energy domain is currently of particular importance for not only the nuclear physics but also some interdisciplinary fields. In the present work, neutron-rich $^{14}$C and $^{16}$C ion beams at 27.5 MeV/nucleon were used to bombard carbon and polyethylene (CD$_{2}$)$_{n}$ targets. Energy and angular distributions of the produced fragments were measured. Background events originating from the carbon content in (CD$_{2}$)$_{n}$ target were efficiently excluded using an extended $E-P$ plot method. Experimental results are systematically analyzed by using HIPSE-SIMON dynamic model. The comparison reveals that, for the carbon target, the HIPSE-SIMON calculation overestimates the yields of the beam-velocity component for fragments near the projectile and also the energy phase space for fragments far away from the projectile, suggesting fine tuning of the overall interaction profile adopted in the model. In contrast, for reactions with the deuteron target, the model calculation can reasonably reproduce the experimental data. The implication of the fragmentation mechanism to the validity of the invariant mass method, as frequently used to reconstruct the clustering resonant structures in light nuclei, is also discussed.

\end{abstract}



\footnotetext[0]{\hspace*{-3mm}\raisebox{0.3ex}{$\scriptstyle\copyright$}2013
Chinese Physical Society and the Institute of High Energy Physics
of the Chinese Academy of Sciences and the Institute
of Modern Physics of the Chinese Academy of Sciences and IOP Publishing Ltd}%

\end{CJK*}

\begin{multicols}{2}
\section{INTRODUCTION}

Understanding the reaction mechanism has always been one of the crucial tasks when looking at the behavior of interacting nuclear systems ~\cite{YeYL-2025, weik2024, Cook-2023}. At low energies (around the Coulomb barrier), the reaction mechanisms are relatively simple, classified mainly into the direct reaction (inelastic scattering, transfer reaction etc. )  and the fusion reaction, for the peripheral and central collisions, respectively ~\cite{Bertulani2004, Cook-2025}. At very high energies (beyond $\sim$100 MeV/nucleon), the collision is dominated by the nucleon-nucleon interactions and can essentially be described by the fully microscopic models ~\cite{Mancusi-2014}. In between of these two extremes is the Fermi-energy region (about 20 to 100 MeV/nucleon), where the movements of nucleons inside the projectile and those of nucleons in the target nucleus can be interwound to create much more complex processes ~\cite{Serber-1947, Lacroix2004}. The typical reaction mechanisms in Fermi-energy range and for heavier nuclei include the deep inelastic scattering, incomplete fusion, multi-fragmentation and others ~\cite{Bertulani2004, Lacroix2004}. 

In recent years, more attention has been attracted to study the nuclear fragmentation reaction owing mostly to the increasing needs for radioactive ion-beam production in many laboratories world wide ~\cite{MaCW-2021} and for the cancer therapy using high-energy heavy-ion (HI) beams ~\cite{Divay-2017}. As a matter of fact, when using high energy carbon-ion beams to treat tumors,  about 50$\%$ of the primary ions undergo nuclear fragmentation and thus may create some unexpected secondary biologic effects (~\cite{Divay-2017, Mei-2023} and references therein). Similar studies have also been required for space science where the galactic cosmic ray radiation is in question ~\cite{Adamczyk-2012}.  Therefore, well validated reaction models for a wide energy range is essential in order to reliably simulate the practical complex processes. In addition, the breakup (fragmentation) of projectiles in Fermi-energy range has frequently been used to investigate the exotic clustering structures, such as chain-like or BEC-like structures , via the invariant mass (IM) reconstruction method ~\cite{Yangzh2023, Chenjh2023, YangZaihong2014PRL, Liuy2020, HanJX-2022, Hanjx2023, ChenY-2025, YuHanzhou2021, Zang_2018, Fritsch2016, Yamaguchi-2017}.  It is also important to clarify the reaction-decay mechanism to avoid the misinterpretation of the resonant peaks. 

In terms of theoretical description of the fragmentation reaction, the most simple way is to use the empirical (parameterized) formulas. The typical ones include EPAX series ~\cite{Summerer-2012} which are adopted in LISE++ code ~\cite{Tarasov-2003} and FRACS series with extension to cover very neutron-rich fragments ~\cite{Mei-2017, Mei-2023, WeiXB-2025}.        
However, in order to understand the underlying physics associated with the observed phenomena and to establish the prediction power onto a firm ground, it is indispensable to build the dynamic model being able to describe the spatial and timing dependence of the reaction processes. The first idea regarding to the fragmentation mechanism came from Serber's supposition ~\cite{Serber-1947}. He compared the velocity of the relatively high-energy projectile with that of the nucleons inside the target nucleus and the possibly transferred nucleons, and suspect a two-step scenario. The primary step (abrasion step) is featured by the fast direct interaction creating an interaction zone and forming the possibly excited clustering species. The secondary step (ablation step) is characterized by the much slower particle evaporation from the remnants of the primary step. This abrasion-ablation (AA) approach certainly depends on the corresponding impact parameter and hence is a real dynamic treatment. The AA approach was later on implemented into calculation codes, such as NUCFRG3 ~\cite{Adamczyk-2012} and HIPSE-SIMON ~\cite{Lacroix2004}. The later is a well validated semi-microscopic dynamic model ~\cite{Mocko-2008, Vient-2018, Frosin2023}, which will be introduced in more details in the following section and used in the present work to interpret the experimental results. Of course there exist some fully microscopic models, of them the most representative one would be the antisymmetrized molecular dynamics (AMD) approach (~\cite{Ono-1999} and references therein). This model grounds the time dependent nucleus-nucleus collision on the nucleon degree of freedom, and, therefore, can be used to assess the direct interaction processes handled in other models. However, the formation of heavier clusters (fragments) , particularly the fusion remnants, in AMD model is still in a trial stage ~\cite{Frosin2023, HanR-2020}.   

In this article, we report on the experimental results of the fragmentation induced by $\rm{^{16}C}$ and $\rm{^{14}C}$ beams at 27.5 MeV/nucleon on carbon and deuterium targets. These results offer additions to the rather sparse fragmentation-reaction data in Fermi-energy domain and for relatively light systems including the typical neutron-rich projectiles ~\cite{Frosin2023, Baldesi2024}. Energy and angular distributions of the detected fragments are compared to the dynamic model calculations using the HIPSE-SIMON code. The understanding of the reaction mechanisms and the the implication to the reconstruction of the clustering resonant states are discussed. Possible improvements of the dynamic model are suggested.    

\section{DESCRIPTION OF THE EXPERIMENT AND THE THEORETICAL MODEL}

\subsection{Experimental setup and performance}

The experiment was performed at the Radioactive Ion Beam Line at the Heavy Ion Research Facility in Lanzhou (HIRFL-RIBLL1) ~\cite{Sun2003}. The radioactive ion beam (RIB) composed primarily of $^{16}$C or $^{14}$C isotope was produced from a 60 MeV/nucleon $^{18}$O$^{8+}$ primary beam impinging on a 3500-$\rm{\mu m}$-thick $^{9}$Be primary target and separated by the RIBLLl beam line using the standard technique ~\cite{ChenY-2025, YangZaihong2014PRL, Liuy2020, Hanjx2023}.  The RIB at an energy of about 27.5 MeV/nucleon and with an average intensity of about 3$\times$$10^{4}$ particles per second (pps) , was tracked by three parallel plate avalanche chambers (PPACs) before impinging on a 31.53-mg/cm$^2$ $^{\rm{nat}}$C target or a 9.53-mg/cm$^2$ $(\rm{CD}_2)_n$ target. The fragments produced from the reaction were detected by a state-of-the-art charged particle telescope (T0) composed of three layers of double sided silicon-strip detectors (DSSDs, labeled D1, D2 and D3) plus three layers of large size silicon detectors (SSDs, labeled S1, S2 and S3) backed by an array of 2 $\times$ 2 CsI(Tl) scintillator array (Fig.\ref{setup}). The angular coverage of T0 is about 0 to 20 degrees. Another array of  Annular Double-sided Silicon Strip Detectors (ADSSDs) was installed around T0, covering the scattering angle of about 26.4 to 56.3 degrees and to be used to detect the recoil light charged particles ~\cite{GeHY-2025}. The details of the detector configurations and performances can be found in ~\cite{ChenY-2025, Zhuhy2023, Lig2021}. In this experiment, a combination of the traditional VME system and also the XIA digital system was adopted for the data acquisition which allows to accept up to 600 signal channels ~\cite{Puwl-2024}. 

We note that a problem occurred during the measurement related to the $^{14}$C beam. The application of the 0-degree telescope is of special advantages of detecting the reaction products at very small angles with respective to the beam direction, which is of particular importance for reactions in inverse kinematics and for the study of clustering resonant states close to the corresponding cluster-separation thresholds ~\cite{Liuy2020, Hanjx2023, Zang_2018}. However, detection in this manner may lead to strong damages of the silicon detector layer which happens to stop the incident beam with high intensity. During the present experiment, this kind of damage occurred evidently to the T0-D2 detector when $^{14}$C beam was impinged. Fortunately, the damage was concentrated at the central area within a radius of about 20 mm of the D2 detector whereas the outer area (more than half of the detector active size) could still be used at least for the heavier fragments (beyond beryllium). The effect of this cutoff on D2 layer, and consequently on T0 telescope, will be discussed in the follow sections. 

Using the self-uniform calibration method ~\cite{QiaoR2014}, energy match for different strips in one DSSD was achieved. Then the energy-loss matching between the hits on both sides of one DSSD and the tracking matching between hits from different layers of DSSDs were employed to assign hits to certain particles traveling the multilayer DSSDs in T0 telescope. Finally it was required that the assigned particle must belong to one of the particle identification (PID) bands defined by the energy loss ($\Delta E$) in one silicon layer versus the remaining energy ($E$) measured by the subsequent detector layer ~\cite{ChenY-2025, Liuy2020, Hanjx2023}.  Fig.\ref{16C+C PID} shows one example of the PID spectra which exhibits clear discrimination of the detected isotopes. The absolute energy calibration of the detectors was realized by using the radioactive $\alpha$-particle sources and also by comparing the experimental PID curve with those calculated using energy-loss tables ~\cite{Ziegler-2010} . This method was validated by using the two- and three-$\alpha$ coincident events to reconstruct the known $\rm{^8Be}$ and $\rm{^{12}C}$ resonant states ~\cite{LiuYC-2023, Hanjx2023}. 

\begin{center}
    \includegraphics[width=235pt]{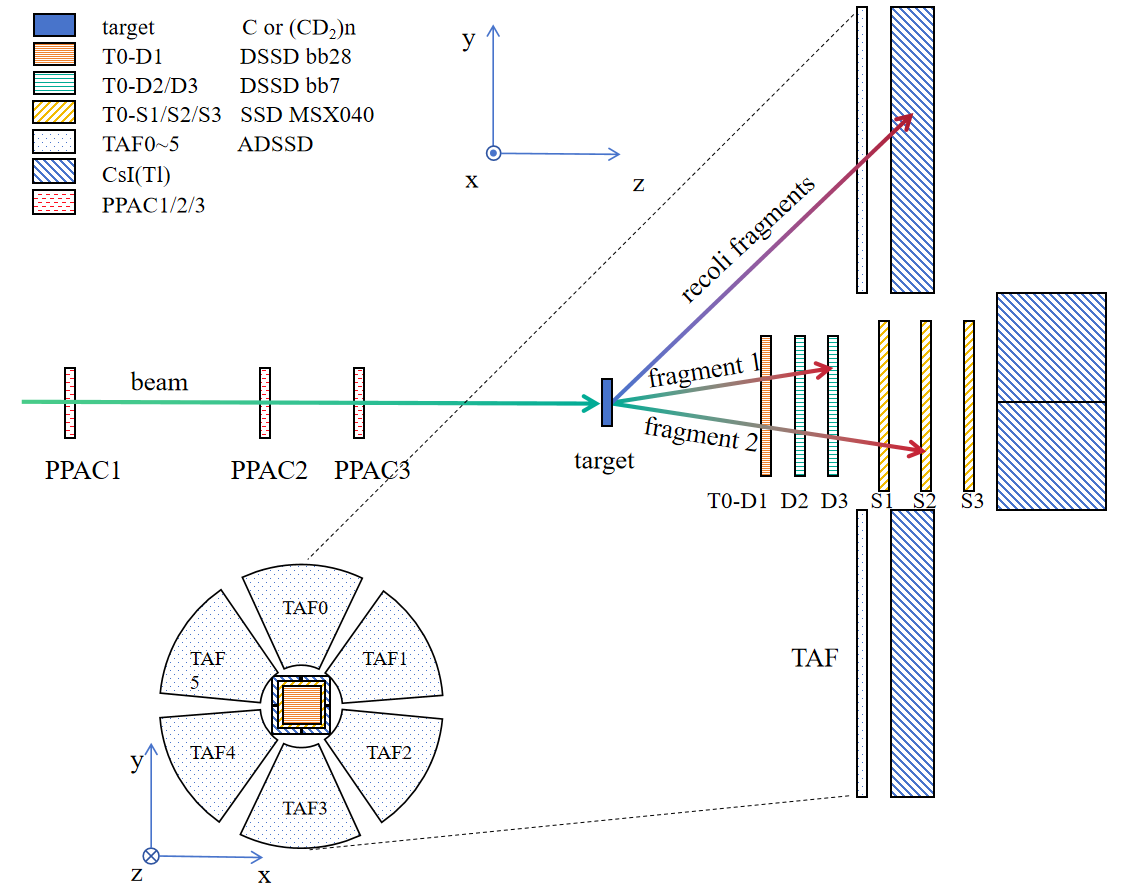}
\begin{figure}
        \centering
        \label{fig:enter-label}
    \end{figure}
        \figcaption{\label{setup} Schematic view of the experimental setup. The beam is tracked by PPACs onto the target. Reaction fragments are detected by the T0 telescope, consisting of DSSDs and SSDs, and the surrounding TAF array. See texts for details.}
\end{center}

\begin{center}
    \includegraphics[width=235pt]{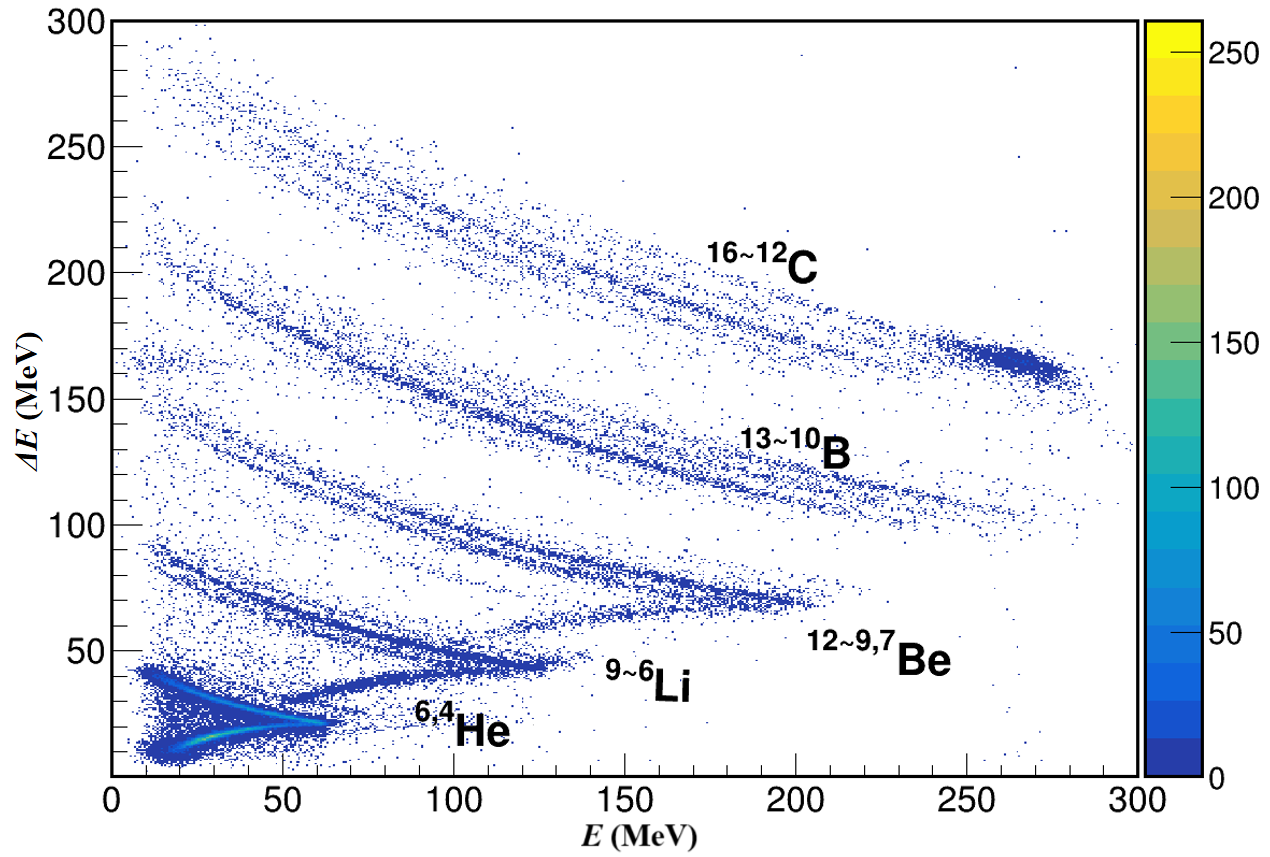}
\begin{figure}
        \centering
        \label{fig:enter-label}
    \end{figure}
        \figcaption{\label{16C+C PID}  {Example of Particle identification (PID) spectra defined by the energy loss ($\Delta E$) and the remaining energy ($E$) as measured by T0-D1 and T0-D2 detectors, respectively, for the $^{16}$C + C reaction at 27.5 MeV/nucleon. Each band corresponds to a specific isotope, demonstrating clear particle identification. For each element, mass numbers of the isotopes are labeled, with the higher number corresponding to the upper band. }}  
\end{center}

\subsection{Dynamic reaction model HIPSE-SIMON}

Heavy ion phase space exploration (HIPSE) model ~\cite{Lacroix2004} was implemented to bridge the gap between the empirical parameterized models, such as EPAX  ~\cite{Summerer-2012} and FRACS ~\cite{Mei-2017}, and fully microscopic models, such as QMD ~\cite{MaCW-2021} and AMD ~\cite{Ono-1999}. It takes a macroscopic way to handle the overall movement of the colliding partners and fragments, but treat the reaction participants and the cluster formation on a microscopic footing. It has also made a realistic allocation of the excited energies for the established fragments which then go through the secondary decay step according to the statistical method. 

HIPSE model was essentially configured into three phases ~\cite{Lacroix2004}. 

(a) The approaching phase of the collision. This phase follows the Classical two-body dynamics of the center of masses. The interaction potential is essentially the Coulomb barrier at large distance while the nuclear potential becomes effective at small relative distance which can be tuned by a phenomenological parameter ($\alpha _a$). At the minimal distance of approach, nucleons in colliding nuclei are sampled based on a zero-temperature Thomas-Fermi distribution, corresponding to the frozen density limit. Then the participant
region is defined by a geometrical evaluation of the overlapped nucleons. Nucleons outside the overlap region give the quasi-projectile (QP) and quasi-target (QT) spectators. Here two local physical processes, namely direct nucleon-nucleon collisions and nucleon exchange, are treated by hand. At higher energies, direct nucleon-nucleon (NN) collisions becomes increasingly important, which is handled by assuming a fraction, defined by the parameter $x_{\rm{coll}}$, of the nucleons in the overlap region undergoes in-medium NN collisions. Another local process is to have a fraction, defined by the parameter $x_{\rm{ex}}$, of the nucleons, which are within the overlap region and originally linked with one spectator, being transfer into another spectator. These two local processes would relax the pure participant-spectator picture and to form more sophisticated partitions (see below). 

(b) The partition formation phase. This corresponds to the rearrangement of the nucleons into fragments, including clusters and light particles, according to the impact parameter of the reaction. The fragments are built following coalescence algorithm based on the evaluation of the relative kinetic and potential energies among all nucleons. The highly fragmented species from this stage would keep a strong memory of the entrance channel. At this point, time $t = 0 {~\rm{fm}}/c$ is set for the forthcoming dynamics.

c) Final states interaction and the reaggregation phase.  To account for the final state interactions (FSI) between the initial fragments, which is particularly significant in the dense matter environment, the clusters are propagated according to a classical Hamiltonian by using the same nucleus-nucleus potential as in the approaching phase. The reaggregation is incorporated in this phase to allow for combination of some of the initial fragments according to the final state interaction, including fusion processes. At the end of this stage, the clusters can no longer exchange particles and a chemical freeze-out is reached. The total excitation energy is then determined from the energy conservation and assigned to each cluster. 

All above three phases constitute the primary step of the reaction, which normally last for about 50 fm/c. This time period is set to allow the reaction-produced fragments to flight away from the nuclear interaction zone and to reach a chemical freeze-out before getting into the much slower decay process. The phase space generated by HIPSE can then be feed into another code which handles the secondary in-fight decay according to the well established statistical method. We adopt the SIMON ~\cite{Durand-1992} code which has usually been combined with HIPSE. HIPSE+SIMON model has long been used to describe fragmentation reactions in Fermi energy domain and often been compared with AMD calculations ~\cite{Mocko-2008, Vient-2018, Frosin2023}. It has advantages of fast running and can provide reasonable insight into the timing and spacial dependence of the dynamic processes. It was demonstrated to be able to give comparable results as those from AMD calculations for most QP or QT products and even better for largely altered products such as the fusion remnants ~\cite{Frosin2023}.         

\section{EXPERIMENTAL RESULTS AND ANALYSIS}

We present here the energy spectra for fragments produced in the present experiment which are compared with the corresponding HIPSE-SIMON model calculations. The main purpose is to see the applicability of the model for neutron-rich projectiles interacting with light targets.  

\subsection{Fragment-energy spectra for $^{16,14}$C + C reactions }

Although natural carbon target ($\sim 99\%$ $^{12}$C and $\sim 1\%$ $^{13}$C) was employed, the effect of $^{13}$C constituent can be ignored due to its tiny proportion and non-peculiar target property as compared to $^{12}$C ~\cite{Divay-2017}. Even for $^{13}$C fragment, the detection at forward angles select essentially the projectile-like component, while the recoil target component would predominantly be deviate to vary large angles or even be stopped in the target material. Therefore, in the present work we will simply consider $^{\rm{nat}}$C as a pure $^{12}$C target. Fig.\ref{16C+C_EperA} shows the energy spectra for fragments produced in $^{16}$C + C reaction at an incident beam energy of 27.5 MeV/nucleon. As depicted in the PID spectrum (Fig.\ref{16C+C PID}), the well-identified fragments include the C, B, Be and He isotopes. In order to reduce the background in PID spectrum induced by the extremely intense beam particles (namely $^{16}$C), a requirement was applied to reject particles with scattering angle smaller than 3.6 degrees as selected from the scattering angle spectrum. Even with this requirement, it is still difficult to obtain clear $^{15}$C energy spectrum due to some contamination at its higher energy tail. Thus, in Fig.\ref{16C+C_EperA} we plot the experimental data starting from $^{14}$C fragment. Similar to carbon isotopes, boron fragments were also stopped by the first two layers (D1 and D2) of T0 telescope and the energy spectra can be directly extracted as depicted in Fig.\ref{16C+C_EperA} ($^{13,12,11,10}$B in 2nd row). For beryllium fragments, the penetration to the third layer (D3) of T0 telescope happens for some higher energy species, as can be seen from the back-bending behavior of the Be-PID bands. Since the existence of a thin dead film at each surface of the silicon detector, the energy spectrum may show some discontinuities (kinks) at the corresponding energies, as exhibited in $^{12,11,10,9}$Be energy spectra (3rd row in Fig.\ref{16C+C_EperA}) around 28 MeV/nucleon. As for lithium fragments, the penetration into the fourth layer becomes possible and the same kind of discontinuities appear in the spectra (4th row in Fig.\ref{16C+C_EperA}). The helium isotopes may further penetrate into all six layers of silicon detectors and even into the CsI(Tl) crystals of the telescope. In this case, only $^4$He energy spectrum was practically extracted as displayed in the upper-right panel of Fig.\ref{16C+C_EperA}. We note that the data points at the lower energy side of some spectra, such as those for $^{13}$B and $^{9,8}$Li, may have larger uncertainty due to the stronger background for signals produced near the front surface of D2 detector, as can be seen in Fig.\ref{16C+C PID}. From all panels in Fig.\ref{16C+C_EperA}, we see a clear trend of expanding the energy phase space from heavier fragments (closer to the projectile) to lighter isotopes (further away from the projectile), being reminiscent of the transition from quasi-direct reaction processes to more dissipative (violent) reaction processes ~\cite{Frosin2023, Baldesi2024}. Particularly, for light lithium and helium isotopes the energy per nucleon may largely excess the beam energy per nucleon. 

\end{multicols}
\begin{center} 
    \includegraphics[width=450pt]{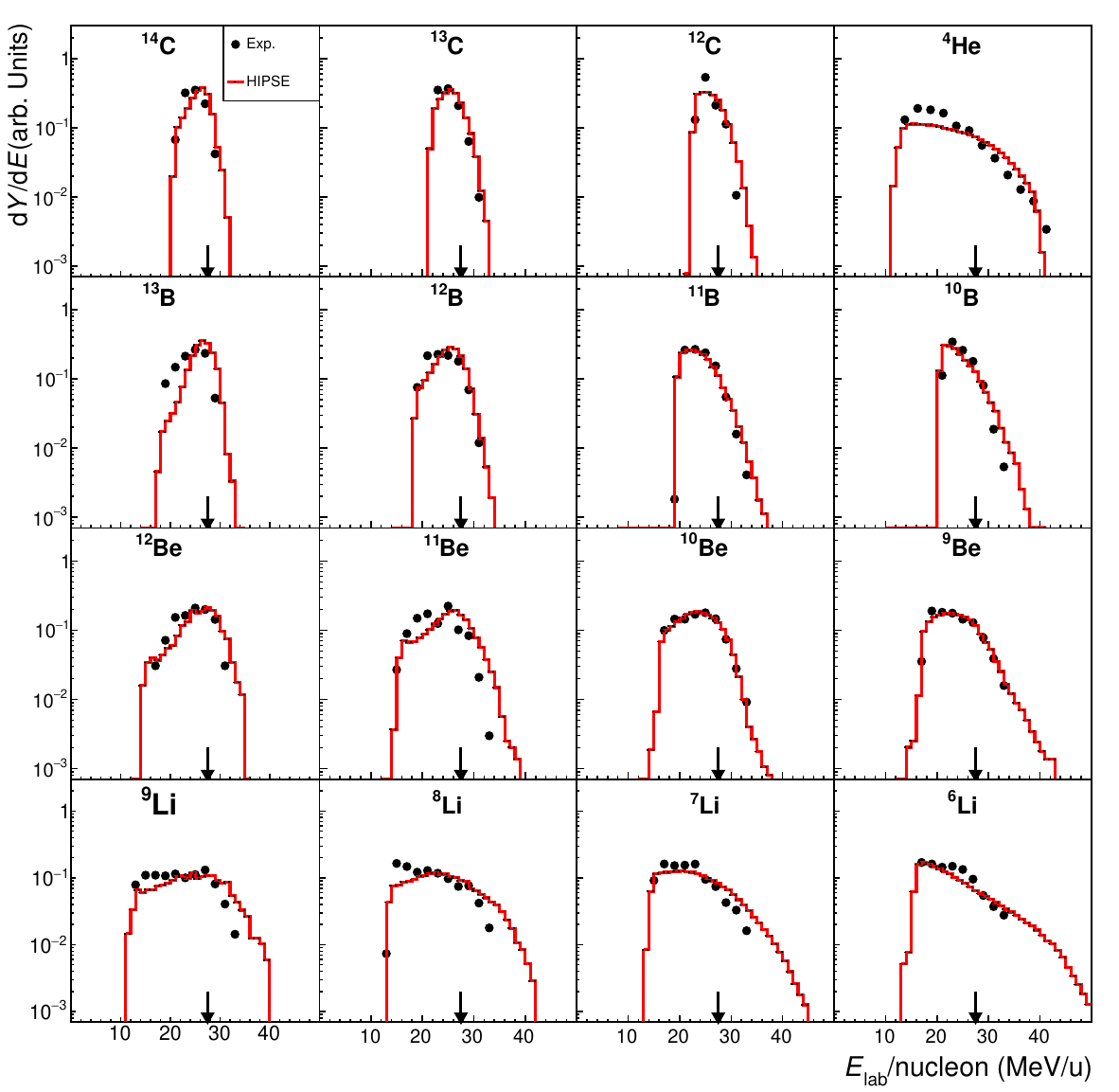}
    \figcaption{\label{16C+C_EperA} Experimental Energy spectra (black-solid points) in comparison to the corresponding HIPSE-SIMON model calculations (red-solid lines) for fragments produced in the reaction $^{16}$C + C at 27.5 MeV/nucleon ~\cite{Frosin2023, Baldesi2024}. Each spectrum is normalized to its integral for a better shape comparison. The arrow in each panel indicates the incident beam energy per nucleon (``$E_{\rm{beam}}/{\rm{nucleon}}$''). The fragment (isotope) names are labeled in respective panels. The error bars are mostly small compared with the data-point size in the current logarithmic presentation and hence omitted in the figure.}
\vspace{2mm}
\end{center}
\begin{multicols}{2}

Similar experimental results were also obtained for $^{14}$C + C reaction and the fragment-energy spectra are given in Fig.\ref{14C+C_EperA}. As noted in previous section, the central part of the T0 telescope has to be excluded from the $^{14}$C-beam related data analysis and hence the extracted spectra correspond to larger angle emission associated with more violent reaction processes. Moreover, we did not accept the spectra for lithium and helium isotopes here because of the contamination of the PID spectrum, associated with the intense $\alpha$-particles hitting the damaged D2 detector.  Again, the expansion of the energy phase space from heavier fragments to lighter isotopes is exhibited in the spectra (Fig.\ref{14C+C_EperA}).   

\end{multicols}
\begin{center} 
    \includegraphics[width=450pt]{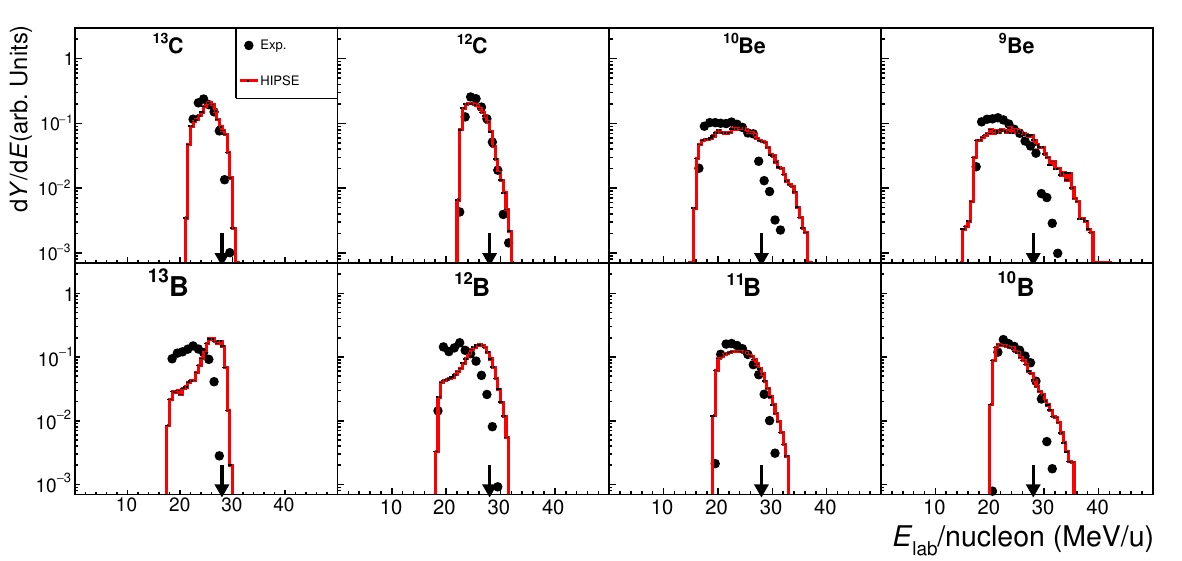}
    \figcaption{\label{14C+C_EperA} Similar to Fig.\ref{16C+C_EperA} but for $^{14}$C + C reaction at 27.5 MeV/nucleon.}
\vspace{2mm}
\end{center}
\begin{multicols}{2}

In Fig.\ref{16C+C_EperA} and \ref{14C+C_EperA}, we present also the HIPSE+SIMON model calculations for comparison. The conventional model-parameter values suitable to the present energy range ~\cite{Lacroix2004}, namely $\alpha_0 = 0.1$, $X_{\rm{tr}} = 0.45$ and $X_{\rm{coll}} = 0.01$, are adopted. Some changes on these values were tested but no visible effect was found regarding to the investigated spectrum shapes. Of course the model generated fragments, together with their energy and momentum, must be processed according to the realistically simulated detector configurations and data-analysis conditions ~\cite{Hanjx2023, ChenY-2025}, before being used in comparison with experimental data. From Fig.\ref{16C+C_EperA} and \ref{14C+C_EperA}, we see that the model calculations reproduce quite well the general form of the experimental spectra for both $^{16}$C and $^{14}$C projectiles. However, two systematic discrepancies between the calculations and the data appear in the figures. One is the model overestimation of the beam-velocity ($E_{\rm{beam}}/{\rm{nucleon}}$) component for fragments close to the projectile in mass, such as $^{14}$C and $^{13,12}$B isotopes from $^{16}$C + C reaction (Fig.\ref{16C+C_EperA}), and $^{13}$C and $^{13,12}$B isotopes from $^{14}$C + C reaction (Fig.\ref{14C+C_EperA}). Another is the model overestimation of the higher energy tail for fragments far away from the projectile, such as $^{8,7,6}$Li from $^{16}$C + C reaction (Fig.\ref{16C+C_EperA}), and $^{12,11,10}$B and $^{10,9}$Be isotopes from $^{14}$C + C reaction (Fig.\ref{14C+C_EperA}). We note that in the case of $^{14}$C beam, the displayed experimental or calculated spectra , are in favor of the more dissipative processes due to the cut off of the central part of the T0 telescope as remarked in above section. These two discrepancies will be further addressed in the following discussion section.    

\subsection{Fragment-energy spectra for $^{16,14}$C + $d$ reaction} 

In recent years, there is an increasing tendency to use simple-structured light targets, such as proton and deuteron, to probe the exotic structure of unstable nuclei being provided as the incident beam particle ~\cite{CaoZX-2012, SunXH-2020, LiuHN-2022, Santamaria-2018, LiPJ-2023, Matsuda-2011, Liuy2020, HanJX-2022, Hanjx2023, ChenY-2025}. However, quite often composite materials, namely $(\rm{CD}_2)_n$ or $(\rm{CH}_2)_n$, are used instead of pure proton or deuteron targets. Since the fragment counting rate resulted from the deuteron component of the target is much lower than that from the carbon component, it would be  difficult to extract results for deuteron target by a direct subtraction method ~\cite{Divay-2017}. In the present work, we apply an extended ``$E-P$ plot'' method, which allow to select reaction products primarily associated with the deuteron component of the target through the kinematic conditions.  

\subsubsection{Extended $E-P$ plot method to select the target component}

For a reaction-decay process in inverse kinematics, namely $\rm{A(a,b)B^*\rightarrow c + d}$ with ``A'' the projectile, ``a'' the target nucleus, ``b'' the recoil target-like particle and ``B$^*$'' the reaction produced nucleus in a resonant state. By measuring coincidentally the decay fragments ``c'' and ``d'' and by applying the energy-momentum conservation, the mass of the target nucleus can be determined from the slope of a band in the ``$E-P$ plot'' spectrum ~\cite{weik2024, Costanzo1990}. Now if we measure the recoil particle ``b'' (deuteron in the present case) at large angles in coincidence with one of the decay fragment (``c'' for instance), the energy-momentum conservation leads to         
 \begin{equation}\label{EPplot1}
    \begin{split}
      \boldsymbol{p}_{\rm{d}} &= \boldsymbol{p}_{\rm{A}} - \boldsymbol{p}_{\rm{c}} - \boldsymbol{p}_{\rm{b}}, \\
       T_{\rm{d}} &= T_{\rm{A}} - T_{\rm{c}} - T_{\rm{b}} + Q .
     \end{split}
    \end{equation}
Defining  
 \begin{equation}\label{EPplot2}
    \begin{split}
        X &= \boldsymbol{p}^2_{\rm{d}} /2u_0  ,\\
        Y &=  T_{\rm{d}} - Q = \frac{X}{A_{\rm{d}}} - Q    
    \end{split}
    \end{equation}
with $u_0$ the nucleon mass. The two dimensional plot on $X$ versus $Y$ (``$E-P$ plot'') may exhibit a band with a slope $1/A_{\rm{d}}$ if ``d'' represent a well defined nucleus which must uniquely related to the mass of the actual target component ~\cite{weik2024}.
\end{multicols}
\begin{center}
    \includegraphics[width=350pt]{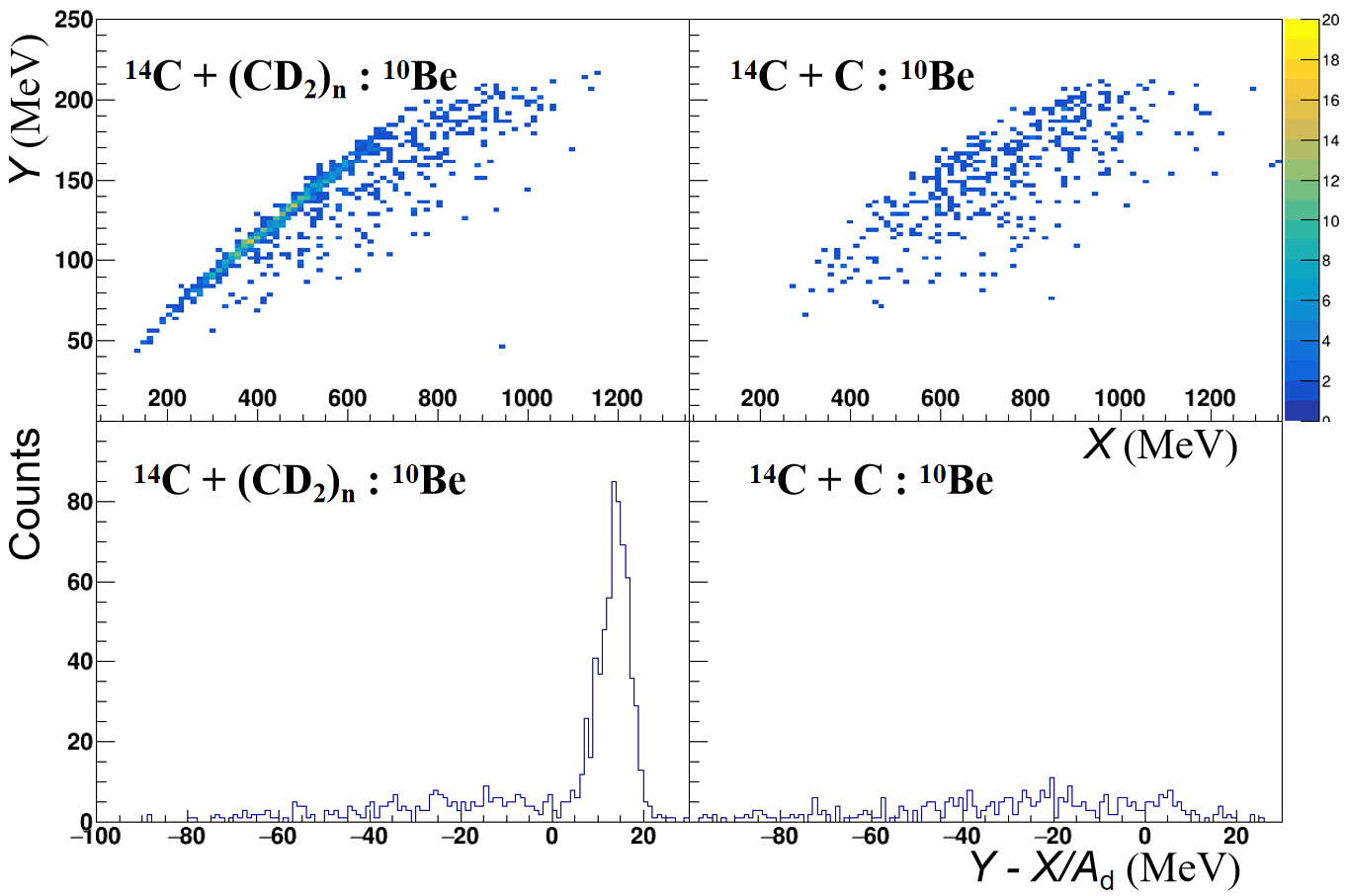}
\begin{figure}
        \centering
        \label{fig:enter-label}
    \end{figure}
        \figcaption{\label{EPplot} {An example of applying the extended $E-P$ plot method, using data for coincidentally measured $^{10}$Be fragment and recoil deuteron, produced in the reaction induced by $^{14}$C beam at 27.5 MeV/nucleon on (CD$_2)_n$ and carbon targets. The upper two panels show  two-dimensional spectra for $Y$ versus $X$, while the lower two panels present their projections, respectively, along the direction of the 1/$A_{\rm{d}}$ slope. $X$, $Y$ and $A_{\rm{d}}$ are defined in Eq.\ref{EPplot2}}.}
\end{center}
\begin{multicols}{2}
\par Taking our case as an example (Fig.\ref{EPplot}), by measuring the emitted recoil deuteron and one of the beryllium fragments ($^{10}$Be for instance) induced by $^{14}$C beam on deuteron target, the extracted band-slope parameter should correspond to $^4$He isotope (see the spectrun in the upper left panel of Fig.\ref{EPplot}). However, if the reaction occurs on the carbon target, the same measurement would results in much different slope parameter or even scattered distribution in $E-P$ plot spectrum (see the upper right panel of Fig.\ref{EPplot}). We note that even for $^{14, 16}$C + C reaction, deuterons can still be emitted and detected from processes such as projectile breakup or fusion-evaporation and so on, and could therefore contaminate the intended process with the deuteron target. By applying a gate on the relevant peak in the projected one-dimensional spectrum (see an example in the lower left panel of Fig.\ref{EPplot}), the actual extended $E-P$ plot method can eliminate most of C-target contamination for the heavier fragments (Be, B and C) because of the possibly well formed particle ``d''. The validation of the extended $E-P$ plot method was checked by comparing to the results for pure C-target (such as the spectrum in the lower right panel of Fig.\ref{EPplot}). In the following analysis a contamination from C-target of less than $\sim 10\%$ was required.     

\subsubsection{Experimental results for deuteron target}

As explained above, we now need to measure one of the breakup fragments detected by T0 telescope in coincidence with the recoil deuteron detected by the large-angle array of ADSSDs. According to the reaction kinematics and the experimental setup, only the higher energy branch of the deuterons can be detected. The angular coverage of the deuteron detectors (ADSSDs) also leads to reducing the number of detected fragments at small angles and, hence, the effect of the direct beam. In Fig.\ref{16C+D_EperA} we present the fragment-energy spectra for reaction $^{16}$C + $d$. Here, $^{15}$C fragment can be extracted, in addition to $^{14-12}$C, owing to less beam contamination. However, all spectra for Li and He isotopes and some spectra for B and Be isotopes are not presented due to higher contamination from the carbon target. Similar results are presented in Fig.\ref{14C+D_EperA} for reaction $^{14}$C + $d$. 

HIPSE + SIMON calculations are also compared to the data. Here we see much better agreement  between data and calculations, and also much less expanded energy phase space of the fragments, in comparison to the cases for carbon target. These could be attributed to the weaker interaction (dissipation) induced by deuteron target (refer to further discussion bellow).  

\end{multicols}
\begin{center}
    \includegraphics[width=450pt]{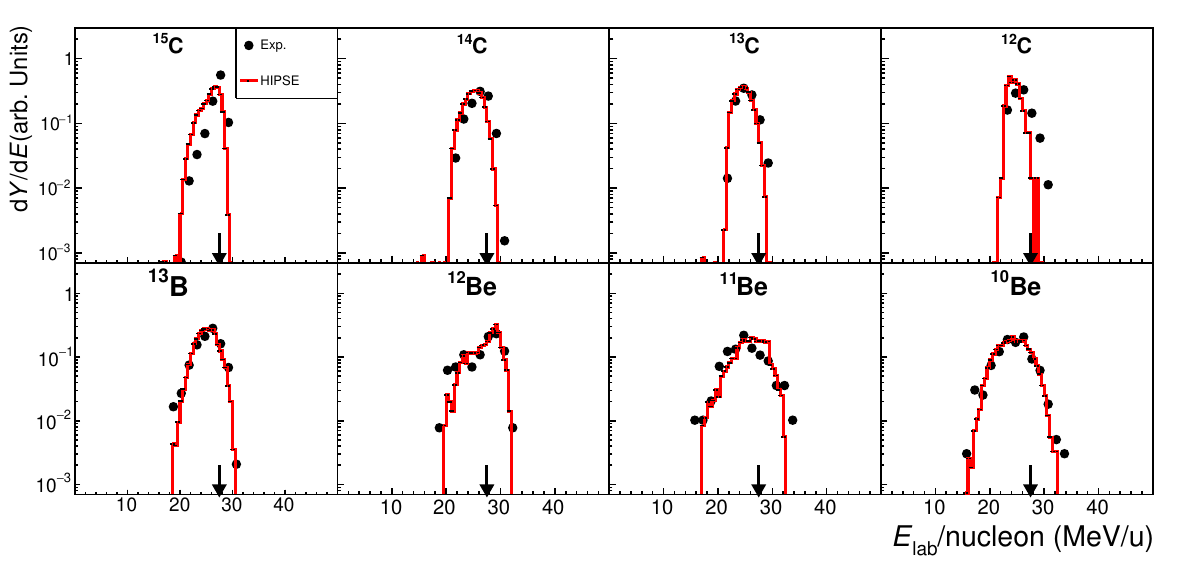}
    \figcaption{\label{16C+D_EperA} Similar to Fig.\ref{16C+C_EperA} but for $^{16}$C + $d$ reaction at 27.5 MeV/nucleon.}
\vspace{2mm}
\end{center}
\begin{multicols}{2}

\end{multicols}
\begin{center}
    \includegraphics[width=350pt]{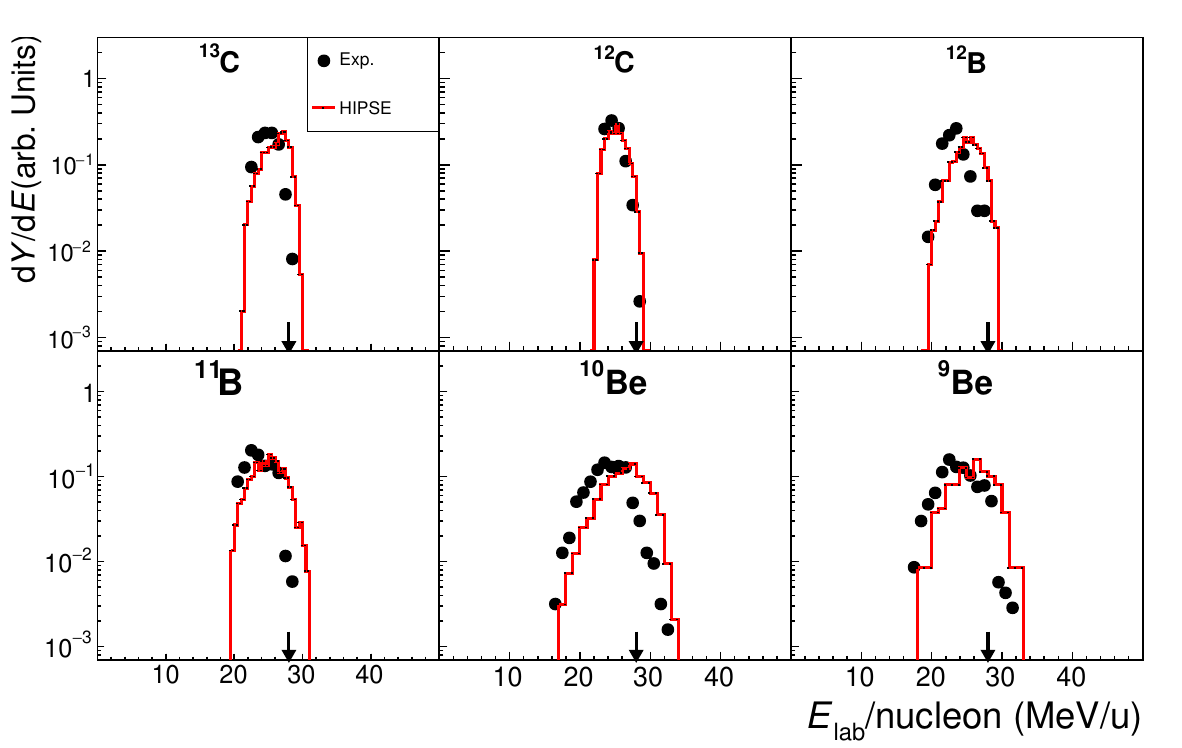}
    \figcaption{\label{14C+D_EperA} Similar to Fig.\ref{16C+C_EperA} but for $^{14}$C + $d$ reaction at 27.5 MeV/nucleon.}
\vspace{2mm}
\end{center}
\begin{multicols}{2}

\section{DISCUSSION}

In the previous section, we have seen some systematic discrepancies between the experimental data and the model calculations for the overall shape of the fragment-energy spectra. Since HIPSE-SIMON is a dynamic model incorporating space (impact parameter) and time (multi-step) dependent processes, it would be possible to infer the dynamic reason of these discrepancies and to suggest possible improvements of the model. The impact of the reaction mechanism to the extraction of the structure information could also be investigated within the model capabilities.  

\subsection{Kinetic-energy spectrum shape}

As indicated in above section, the model overestimates the beam-velocity  component for fragments close to the projectile. In Fig.\ref{16C+C_dbdz} a few representative isotopes produced from $^{16}$C + C reaction are used to illustrate the origin of this component. In the top two panels of Fig.\ref{16C+C_dbdz}, the calculated spectra for $^{14}$C and $^{12}$C fragments are divided into contributions belonging to different impact parameter ($b$) zones. From $^{14}$C to $^{12}$C ($^{13}$C not shown), the contribution of the secondary decay process quickly increases relative to that of the primary reaction process. For $^{14}$C fragment, the primary process is still very significant and hence its dependence on impact parameters can be exhibited. It is evident that both primary and secondary processes contribute to the beam-velocity component, with the former provides even higher energy per nucleon. In middle two panels for $^{13}$B and $^{11}$B fragments, the spectra can be attributed mainly to the secondary decay process summed over contributions from different impact parameter zones. For $^{13}$B we see almost equal contributions from the outer most zone  ($5.2 < b < 6.0 ~{\rm{fm}}$) and the inner zone ($4.2 < b < 5.2 ~{\rm{fm}}$), whereas for $^{11}$B the inner-zone contribution dominates. This change of relative importance of the inner zone can also be seen for the primary process in the top two panels. This tendency is reasonable since the more dissipative inner processes should be in favor of removing more nucleons from the projectile, either in the primary or secondary step. More dissipation for lighter fragments (further away from the projectile) should also be reflected in angular distribution as displayed in Fig.\ref{16C+C_13B11B_T0angle}. It would also be interesting to see the mother nuclei which decay (secondary step) to the actual final isotope. In bottom two panels, we see that $^{13}$B fragment come basically from the same elemental ($Z$ = 5) but heavier isotopes, whereas $^{11}$B come mostly from higher elemental ($Z$ = 7) isotopes. Less contribution from $Z$ = 6 isotopes may be ascribed to the pairing effect which has been included in the model ~\cite{Lacroix2004}. Same analyses have also been performed for $^{14}$C + C reaction and similar phenomena were also identified. In any case, the over estimation of the beam-velocity component for close-projectile fragments, such as $^{14}$C and $^{13}$B here, may be attributed to the insufficient dissipation (overall interaction) in the peripheral area of the reacting nuclei, for both primary and secondary processes. This was casually evidenced in some previous work using stable nucleus beam ~\cite{Baldesi2024} but is systematically demonstrated here for unstable nucleus beam.   

\begin{center}
    \includegraphics[width=245pt]{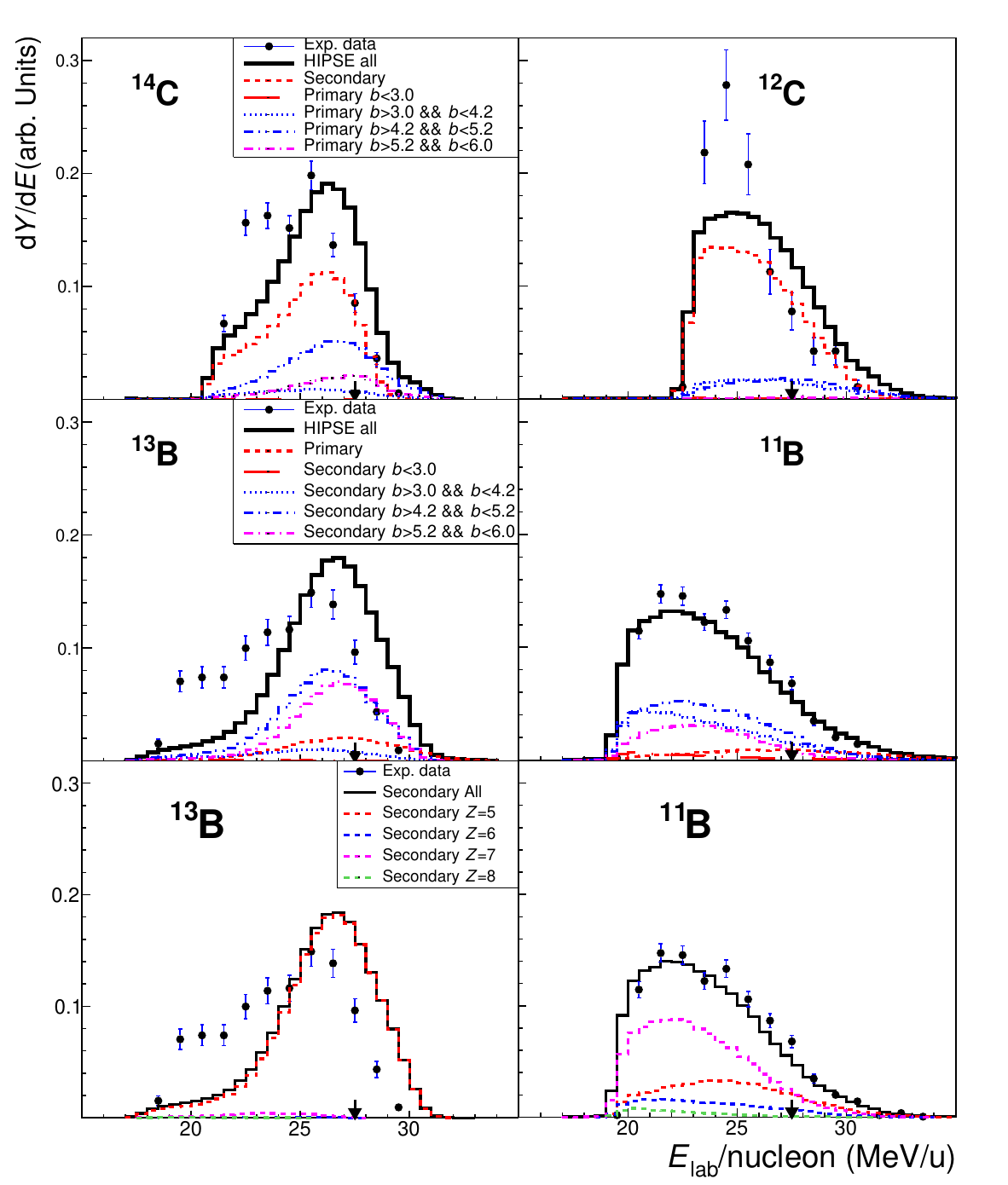}
    \figcaption{\label{16C+C_dbdz} Fragment-energy spectra for selected isotopes produced in $^{16}$C + C reaction. In each panel the calculated spectra are normalized to the integral of the ``all'' spectrum while data points are presented relative the data integral (Error bars are statistical only) ~\cite{Frosin2023, Baldesi2024}. Top panels for $^{14}$C and $^{12}$C isotopes: Experimental data in comparison to HIPSE-SIMON model calculations characterized by primary reaction processes within different impact parameter ($b$, in unit of fm) zones having the equal zone-area; Middle panels for $^{13}$B and $^{11}$B: The comparisons characterized by secondary decay processes within different impact parameter zones; Bottom panels for $^{13}$B and $^{11}$B: Calculated secondary decay processes characterized by mother nuclei which decay to the actual final fragment. Panels in the same row share the same label inset.  }
\vspace{2mm}
\end{center}

\begin{center}
    \includegraphics[width=245pt]{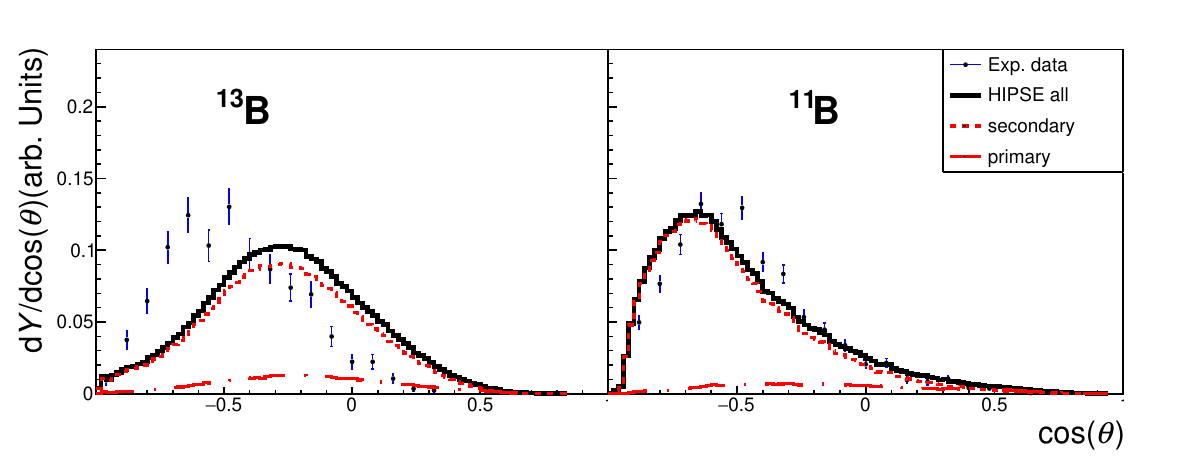}
    \figcaption{\label{16C+C_13B11B_T0angle} Angular distributions of fragments $^{13}$B and $^{11}$B produced in $^{16}$C + C reaction (refer to Fig.\ref{16C+C_dbdz} for the normalized presentation). $\theta$ is the azimuthal angle of the fragment defined in the beam-particle system and with the zero-degree axis pointing to the beam direction. Note that experimental or calculated fragments are filtered by the detection system. Error bars are statistical only. Panels in the same row share the same label inset.   }
\vspace{2mm}
\end{center}

Another problem of the model predictions is the overestimation of the higher energy tails for fragments far away from the projectile. In Fig.\ref{16C+C_7Li_EperA_dbdz} we show a typical example for $^7$Li fragment produced mostly from the secondary decay processes induced by the $^{16}$C + C collision, together with the dynamic analyses associated with impact parameters (left panel) and original mother nuclei (right panel). From the left panel it can be seen that the high energy tail is related to more central collisions ($0 < b < 5.2 ~{\rm{fm}}$). The right panel shows that the actual $^7$Li fragment are originated mainly from the decay of nitrogen isotopes ($Z$ = 7) and even from fusion remnant ($Z$ = 12), again associated most like to the more central collisions. The overestimation of the energy phase space here could then be attributed to too strong the overall interaction within the relatively central part of the collision.                    

\begin{center}
    \includegraphics[width=245pt]{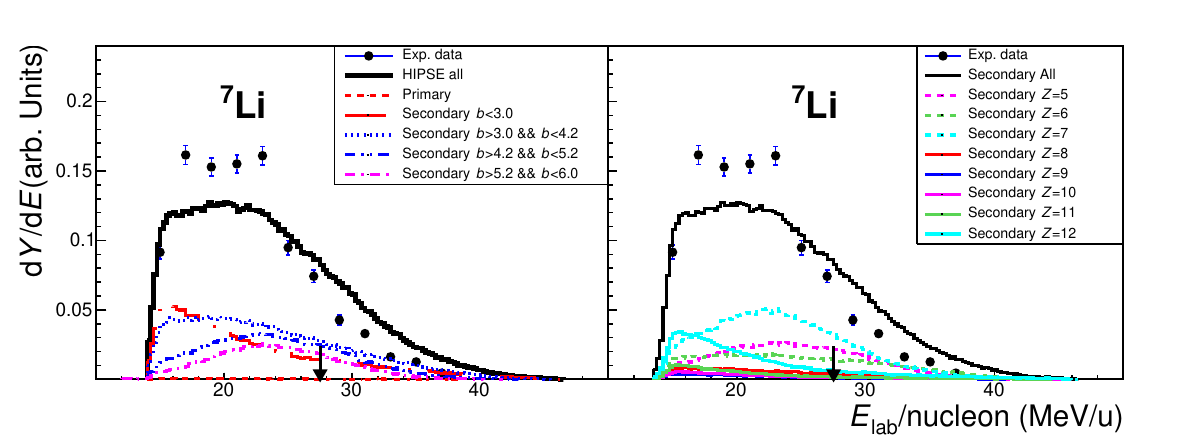}
    \figcaption{\label{16C+C_7Li_EperA_dbdz} Fragment-energy spectra for $^7$Li produced in $^{16}$C + C reaction  (refer to Fig.\ref{16C+C_dbdz} for the normalized presentation). Left panel: Experimental data with statistical error bars are compared with calculations characterized by secondary decay processes within different impact parameter ($b$, in unit of fm) zones. Right panel: Calculated secondary decay processes characterized by mother nuclei which decay to $^7$Li final fragment. }
\vspace{2mm}
\end{center}

Similar dynamic analyses related to impact parameters and two-step processes were also conducted for reactions on deuteron target. Here most of the detected fragments come from the secondary decay processes and the model calculations reasonably reproduce the experimental data. The above investigated problems in model calculations for carbon target seem not evident for deuteron target. This might be attributed to much weaker overall interaction and smaller participation zone during the collision.

\end{multicols}

\begin{table}[t]
\centering
\caption{ Fragment-yields measured in the present experiment for $^{16}$C + C (Exp-C) and $^{16}$C + $d$ (Exp-$d$) reactions,  as relative to the yield of $^{14}$C isotope. Corresponding HIPSE-SIMON model predictions (Cal-C and Cal-$d$, respectively), filtered by the simulated detection and data analysis system, are also listed for comparison.}
\label{Yields}
\begin{tabular}{c|cccccccccccccccc}
\toprule
Frag. & $^{14}$C & $^{13}$C & $^{12}$C & $^{13}$B & $^{12}$B & $^{11}$B & $^{10}$B & $^{12}$Be & $^{11}$Be & $^{10}$Be &  $^{9}$Be & $^{9}$Li & $^{8}$Li & $^{7}$Li & $^{6}$Li & $^{4}$He \\
\midrule
Exp-C     & 1        & 0.56     & 0.22     &  0.67    & 0.79     &  1.73    &   0.29   &  0.15     &   0.27    &  1.82     &    1.10     &   0.33   &  0.75    &  2.83    &  1.00     &  22.1             \\
Cal-C     & 1        & 1.00     & 0.25     &  1.79    & 3.69     &  0.62    &   0.06    &  0.04     &   0.07    &  0.63     &    0.10    &  0.02    &  0.10    &  1.22    &  0.52    &  19.9             \\
Exp-$d$   & 1        & 0.52     & 0.09     &  0.12    &          &          &           &  0.03     &   0.05    &  0.26     &            &          &          &          &          &                   \\   
Cal-$d$   & 1        & 0.48     & 0.01     &  0.33    &          &          &           &  0.04     &   0.08    &  0.25     &            &          &          &          &          &                   \\   
\bottomrule
\end{tabular}
\vspace{5mm}
\end{table}

\begin{multicols}{2}

\subsection{Probabilities of fragment production and nuclear clustering}

The yields of fragments depend on the corresponding production cross sections. In HIPSE-SIMON model, the cross sections are taken from phenomenological formulas which may not be applicable to the present light neutron-rich projectile and light targets. We try to evaluate this by comparing to the present experimental data. Since the detector system covers the major part of the fragment angular distribution and the calculated results can be filtered by the realistically simulated detection and data analysis systems, it is meaningful to use just the measured (or filtered) results.

A Direct comparison between theoretically calculated and experimentally measured absolute yields is not practical here since it requires the precise handling of the target thickness, beam-particle tracking and so on, which may have large uncertainties. To avoid the absolute normalization problem, we may select one isotope very close to the projectile as the reference by assuming  that its absolute cross section can easily be determined from some experiments and can also easily be reproduced by the model calculation. Then we just need to evaluate the actually measured or calculated fragment-yields relative to that of the reference isotope. In the case of $^{16}$C projectile, we choose $^{14}$C-fragment as the reference isotope and the results are listed in Table~\ref{Yields}. For carbon target, the experimental data (first row in Table~\ref{Yields}) show very large production probability for quasi-stable nuclei $^{11}$B, $^{10}$Be, $^7$Li and $^{4}$He, in comparison to that of $^{14}$C. This is consistent with previous measurements using similar beams but at much higher beam energy ( e.g. 240 MeV/nucleon $^{14}$C beam impinging on a carbon target.) ~\cite{Mei-2023}. In terms of model calculation (2nd row in Table~\ref{Yields}), there is a clear trend of overestimation of the yields for heavier fragments but an opposite trend for fragments far away from the projectile, as also visually exhibited in Fig.\ref{Yields-16C-C}. This trend is consistent with the above view about the overall interaction adopted in HIPSE model which seems insufficient for peripheral collision, leading to more survivals of the beam-like nuclei, but too strong for more central collision, leading to stronger energy spread as well as easier breakup of the fragments. In the case of deuteron target (last two rows in Table~\ref{Yields}), the experimentally extracted fragment yields are all smaller than that of $^{14}$C and are mostly similar to those from model calculations, being consistent with the weaker interaction effects as discussed above.            
\begin{center}
    \includegraphics[width=250pt]{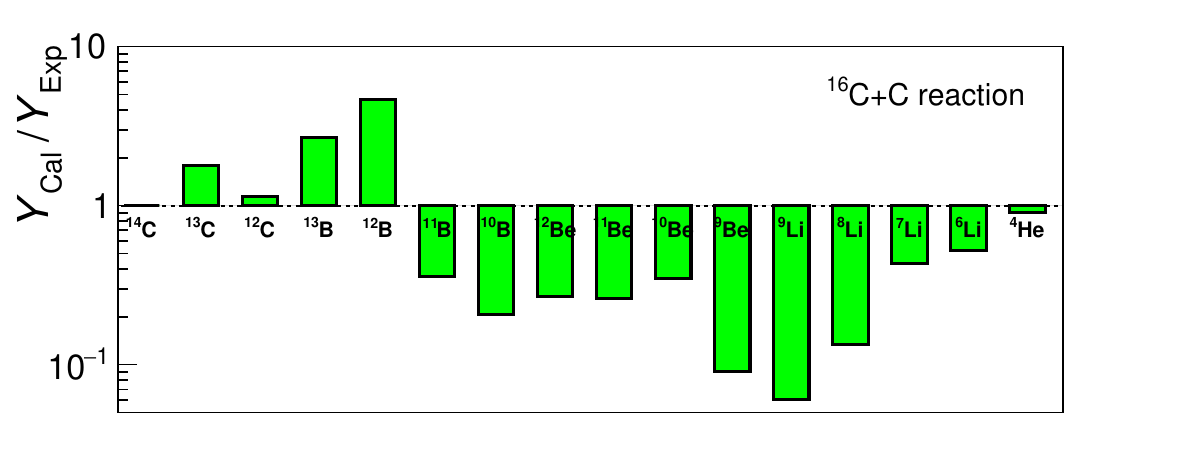}
    \figcaption{\label{Yields-16C-C} Ratio of the calculated fragment yields to the measured ones, produced in $^{16}$C + C reaction. The yield values are taken from Table~\ref{Yields}.}
\end{center}

The general applicability of the HIPSE-SIMON model offers a reasonable insight into the IM method used to reconstruct the clustering resonant states in mother nuclei, as reported in many previous works and summarized in ref.~\cite{weik2024}. Once adopting the two-step scenario for the HI reaction at Fermi-energy range (a few tens of MeV/nucleon),  with the primary step sustaining a short time period of about 50 fm/c ~\cite{Lacroix2004} , the energy spread for any structured source of the primary fragments must be larger than a few MeV according to the uncertainty principal. If these fragments were used to reconstruct the IM of the mother nuclei, the spectrum must look like continuum or very broad configuration. In contrast, if the observed peak in IM spectrum has enough significance and a physical width (excluding the effect of the detection resolution) smaller than about 1 MeV (FWHM), it should come from the secondary decay process with relatively long life-time, reflecting a resonant state. In previous works aiming at studying the clustering resonant states via the IM reconstruction using the emitted fragments ~\cite{YangZaihong2014PRL, Liuy2020, Hanjx2023, ChenY-2025, weik2024}, most reported resonance peaks satisfy this width requirement. What learnt here is that cares should be taken when identifying a very broad structure as a resonant state since it might be contributed from the primary breakup process, not from the secondary decay.         

\section{SUMMARY}

A fragmentation reaction experiment was performed at HIRFL-RIBLL1 facility by using neutron-rich $\rm{^{16}C}$ and $\rm{^{14}C}$ beams at 27.5 MeV/nucleon impinging on carbon and deuterium targets. A state-of-the-art zero degree telescope composed of multi-layer silicon-strip detectors had allowed to measure emitted particles at small angles. Numerous types of fragments were clearly identified together with their energy and angular distributions being measured, providing a good base for systematic analysis. The energy and angular spectra have been compared with HIPSE-SIMON dynamic model calculations. In the case of carbon target, it is recognized that the model overpredicts the beam-velocity component for fragments close to the projectile, likely due to the weakening of the overall interaction at the surface of the colliding nuclei, as used in the model. On the other hand, the calculated energy phase space for fragments far away from the projectile, produced from inner areas of the colliding nuclei, seems too expanded as compared with the experimental data. This might be attributed to the too strong the overall interaction in the central region of the collision. These discrepancies seem mach less evident for deuteron target, due possibly to relatively weak overall interactions. The dynamic analysis also provides credit to the IM method already frequently used to reconstruct the clustering resonant structures from the decay fragments. Present work suggests possible fine tunes of the overall interaction profile used in HIPSE model, particularly for unstable nuclei. One possible way to implement this tuning would be to use more flexible initial nucleon-density distributions of the colliding nuclei. Considering the dynamic feature and the realistic cluster-formation mechanism implemented in HIPSE code, its application to more complicated reaction processes, such as the multi-nucleon transfer reaction, could  be expected. On the experimental side, more measurements with heavier unstable nucleus beams at Fermi energies are badly needed in order to fully understand the nuclear fragmentation mechanism and to support its multidisciplinary applications.     

\vspace{10mm}

\acknowledgments{The authors thank the HIRFL-RIBLL staff for providing excellent technical and operational support during the experiment.}

\vspace{10mm}

\centerline{\rule{80mm}{0.1pt}}


\bibliographystyle{apsrev4-2}

\bibliography{ref}

\end{multicols}
\end{document}